\documentstyle[12pt,a4]{article}

\newcommand{\dslash}{\mbox{$\not{\hspace{-0.8mm}\partial}$}}

\newcommand{\Dslash}{\mbox{$\not{\hspace{-1.1mm}D}$}}
\newcommand{\vslash}{\mbox{$\not{\hspace{-0.8mm}v}$}}

\newcommand{\epsslash}{\mbox{$\not{\hspace{-0.8mm}\epsilon}$}}

\newcommand{\nc}{\newcommand}

\nc{\be}{\begin{equation}}
\nc{\ee}{\end{equation}}
\nc{\bea}{\begin{eqnarray}}
\nc{\eea}{\end{eqnarray}}
\nc{\g}{\gamma}
\nc{\m}{\mu}
\nc{\n}{\nu}
\nc{\r}{\rho}
\nc{\s}{\sigma}
\nc{\e}{\epsilon}
\nc{\k}{\kappa}
\nc{\G}{\Gamma}
\nc{\f}{\phi}

\begin{document}

\begin{titlepage}
\begin{flushright}
 IC/94/64\\
hep-ph/9407330
 \end{flushright}
 \begin{center}
  {\Huge Heavy Quark Effective Theory, Interpolating Fields
 and Bethe-Salpeter Amplitudes}\footnote{Accepted for publication in
Physics Letters B}

 \vspace*{1cm}

 {\large
  F. Hussain and G. Thompson}

 \vspace*{0.5cm}

 {\large
  International Centre for Theoretical Physics, Trieste, Italy\\}

  \vspace*{0.5cm}

{\large May 1994}

  \vspace*{1cm}

 \begin{abstract}
We use the LSZ reduction theorem and interpolating fields, alongwith the
heavy quark effective theory, to investigate
the structure of the Bethe-Salpeter amplitude for heavy hadrons. We show
how a simple form of this amplitude, used extensively in heavy hadron
decay calculations, follows naturally upto $O(1/M)$ from these field
theoretic considerations.

 \end{abstract}

 \end{center}
\end{titlepage}

\section{Introduction}
\label{intro}
\setcounter{footnote}{0}
In recent years the heavy quark effective theory (HQET) \cite{hqet},
\cite{kt} and the consequent
heavy quark symmetry have emerged as useful tools in studying the decay
properties of heavy hadrons. In our previous works \cite{hkstw} -
\cite{htk} we have used the Bethe-Salpeter (B-S) formalism to derive the
consequences of the heavy quark symmetry for weak transitions  of
hadrons of arbitrary spin.

In the heavy quark mass limit we argued
\cite{hkstw}-\cite{htk} that the Bethe-Salpeter (B-S) amplitude ( wave
function ) for an arbitrary heavy meson, of momentum $P$ and mass $M$,
can be written in momentum space as
\be
M_{\alpha}\,^{\beta}(p_{1},p_{2})=
\chi_{\alpha}\,^{\delta}(v,k)A_{\delta}\,^{\beta}(p_{1},p_{2})\,,
\label{bs1}
\ee
where $\chi_{\alpha}\,^{\beta}$ is a projection operator which projects
out a particular spin, parity state from an unknown orbital wave function
$A$. Here $p_{1}$ and $p_{2}$ are the momenta of the heavy and light
quarks respectively and $v$ is the four-velocity of the heavy meson defined as
$v=P/M$. The momentum $k$ is defined as $k=p_{1}-p_{2}$.

The important observation is that $\chi$ has the Bargmann-Wigner \cite{bw}
- \cite{hkt} form
\be
\chi(v,k)=\frac{1}{2}(1+\vslash)\G(k)\,.\label{bw}
\ee
with $\G(k)$, a Dirac matrix, depending on the spin, parity and orbital angular
momentum \cite{htk}. From this simple form of the B-S amplitude follow all the
dramatic results about the reduction of form factors in heavy meson decays.

We argued that this form of the B-S amplitude,
eqs. (\ref{bs1}) and (\ref{bw}), arises because,
in the heavy quark limit,
in the leading order of the heavy quark effective theory (HQET)
\cite{hqet}, \cite{kt} the heavy quark spin is decoupled from the
light degrees of freedom and as a consequence the B-S amplitude
satisfies the Bargmann-Wigner (Dirac) equation on the heavy Dirac index
$\alpha$.
The baryons can also be represented in a similar form.

In this short note we demonstrate that this form of the B-S wave
function follows naturally, upto order $1/M$, from the notions of interpolating
fields and the LSZ reduction theorem, when combined with the zeroth
order HQET. We will demonstrate this
explicitly for the  $0^{-}$ and $1^{-}$ heavy mesons and then show
how it is easily generalised to arbitrary heavy meson resonances and
heavy baryons. As an intermediate step we show that, in the heavy quark
mass limit, only half the
components of the quark fields appearing in the interpolating field
contribute to the matrix element. For example, for a heavy
$(Q\overline{q})$ meson, in the rest frame of the {\em meson} only the quark
part of the heavy quark field \underline{and} only the antiquark part of
the light quark field contribute.

In Section \ref{inter} we discuss the consequences of the
heavy quark mass limit for interpolating fields. In Section \ref{hqet}
we combine the results of Section \ref{inter} with the HQET to obtain
the desired form of the B-S amplitude.

\section{Interpolating Fields and the Heavy Quark Mass Limit}
\label{inter}
As we mentioned in the Introduction, a variety of interpolating
fields can be used
to represent a particular bound state within a reduction formula
\cite{l}. In this section we will first use the equations of motion to
establish
relations between certain classes of interpolating fields and the normalisation
constants
appearing in these interpolating fields. We will then look at these relations
in the heavy quark limit. Then we will study the consequences of these
relations when we use these interpolating fields in reduction theorems.
As a first example consider the pseudoscalar heavy meson for which we can take
as
interpolating fields
\be
\f_{1}(x)=Z_{3}^{1/2}\frac{1}{N_{1}}T\bar{\psi}_{q}(x)\g_{5}\psi_{Q}(x)
\label{ipf1}
\ee
or
\be
\f_{2}(x)=-Z_{3}^{1/2}\frac{1}{N_{2}}
\frac{i\partial_{\m}}{M}T\bar{\psi}_{q}(x)\g^{\m}\g_{5}\psi_{Q}(x)\,,\label{ipf2}
\ee
where the normalisation constants
\be
N_{1}=\langle 0\vert T\bar{\psi}_{q}(0)\g_{5}\psi_{Q}(0)\vert P\rangle
\label{n1}
\ee
and
\be
N_{2}=-\langle
0\vert T\bar{\psi}_{q}(0)\vslash\g_{5}\psi_{Q}(0)\vert P\rangle \label{n2}
\ee
are chosen to ensure the normalisation condition
\be
\langle 0\vert\f_{i}(x)\vert P\rangle=Z_{3}^{1/2}e^{-iP\cdot x}\,,\label{nc}
\ee
for $i=1,2$.

One can now use the equations of motion
\be
(i\Dslash-m)\psi=0 \label{em}
\ee
for the quark fields to show that
\be
\f_{2}=\frac{m_{Q}+m_{q}}{M}\frac{N_{1}}{N_{2}}\f_{1}\label{eq1}
\ee
and, further, from the normalisation condition eq. (\ref{nc}) we get
\be
\frac{N_{1}}{N_{2}}=\frac{M}{m_{Q}+m_{q}}\,, \label{eq2}
\ee
where $m_{Q}$ ($m_{q}$) is the mass of the heavy (light) quark, leading
to $\f_{1}=\f_{2}$. Note that in the heavy quark limit, where
$m_{Q}+m_{q}\rightarrow M$, $N_{1}/N_{2}\rightarrow 1$.

The pseudoscalar case is particularly simple and perhaps not very
enlightening. Much more interesting is the vector meson case which
demonstrates some generic properties of interpolating fields which are
not apparant in the pseudoscalar case.
We will, in fact,  show that the
corresponding two interpolating fields in the vector, $1^{-}$, case are not
exactly equal but differ by a term which is of $O(\frac{1}{M})$ and can
be ignored in the infinite mass limit. It will be then easy to show that
such a situation always occurs for all meson resonances.

An interpolating field
$\f_{\m}(x)$ for a vector particle, with momentum P, should satisfy
\be
\langle 0\vert \f_{\m}(x)\vert P\rangle
=Z_{3}^{1/2}\e_{\m}e^{-iP\cdot x}\,,\label{ncv}
\ee
where the polarisation vector $\e_{\m}$ satisfies the transversality
condition $v\cdot\e=0$ and $\e^{*\m}\e_{\m}=-1$. Thus, as in the
pseudoscalar case, we can consider the following two interpolating fields
\be
\f_{\m}^{1}(x)=Z_{3}^{1/2}\frac{1}{V_{1}}T\bar{\psi}_{q}(x)\g_{\m}^{\bot}\psi_{Q}(x)
\label{ipfv1}
\ee
and
\be
\f_{\m}^{2}(x)
=-Z_{3}^{1/2}\frac{1}{V_{2}}\frac{i\partial_{\n}}{M}T\bar{\psi}_{q}(x)\g^{\n}\g_{\m}^{\bot}\psi_{Q}(x)
\label{ipfv2}
\ee
where $\g_{\m}^{\bot}=\g_{\m}-\vslash v_{\m}$. The normalisation
constants, $V_{i}$, with $i=1,2$, are given by
\be
V_{1}=-\langle 0\vert \bar{\psi}_{q}(0)\epsslash^{*}\psi_{Q}(0)\vert P\rangle
\label{ncv1}
\ee
and
\be
V_{2}=\langle 0\vert \bar{\psi}_{q}(0)\vslash\epsslash^{*}\psi_{Q}(0)\vert
P\rangle
\label{ncv2}
\ee

Now one can use the equations of motion eq. (\ref{em}) for the quark
fields to obtain
\be
i\partial_{\n}\bar{\psi}_{q}(x)\g^{\n}\g_{\m}^{\bot}\psi_{Q}(x)=
-(m_{Q}+m_{q})[\bar{\psi}_{q}(x)\g_{\m}^{\bot}\psi_{Q}(x)-
2i\bar{\psi}_{q}(x)\frac{\vec{D}_{\m}^{\bot}}{m_{Q}+m_{q}}\psi_{Q}(x)]
\label{eqv}\,.
\ee
where the transverse covariant derivative,
$\vec{D}_{\m}^{\bot}=\vec{D}_{\m}-v_{\m}v\cdot \vec{D}$, acts to the right.

One sees that in this case the two interpolating fields are not exactly equal
but differ by a term which can be neglected, as it is small in the heavy
quark limit. Note
that the term neglected, $D_{\m}^{\bot}/M$, (as $m_{Q}+m_{q}\rightarrow M$)
is exactly the kind
of term which is dropped in deriving the HQET at leading order \cite{kt}.
Physically this
means that we consider all momenta transverse to the line of flight of
the meson to be small. One can also look at this result in another way.
$i\bar{\psi}_{q}(x)\frac{\vec{D}_{\m}^{\bot}}{M}\psi_{Q}(x)$ is in fact
another possible candidate for the interpolating field for the vector
meson but its overlap with the physical state becomes negligible in
the heavy mass limit. Note also that
$i\bar{\psi}_{q}(x)\frac{\vec{\partial}_{\m}^{\bot}}{M}\psi_{Q}(x)$
corresponds to a p-wave contribution to the vector, $1^{-}$, state coming
from the anti-quark part of the heavy quark. In the zeroth order HQET
recall that the heavy quark and antiquark are described by separate
fields or to put it another way, the antiquark part of the heavy quark
field is suppressed and thus the overlap of
this interpolating field with the physical heavy meson state becomes
very small in this limit.

Hence, only in
the heavy quark limit, with $m_{Q}+m_{q}\rightarrow M$, the two
interpolating fields, $\f^{1}_{\m}$ and  $\f^{2}_{\m}$ are equal along with
the important relation
\be
V_{1}=V_{2}\,,
\ee
between the normalisation constants.

The above considerations turn out to be a general feature of the heavy
quark limit. Given a particular interpolating field
$N_{1}T\bar{\psi}_{q}{\bf\G}\psi_{Q}$ then one can show that this is
equal to the interpolating field
$N_{2}\frac{-i\partial_{\n}}{M}T\bar{\psi}_{q}\g^{\n}{\bf\G}\psi_{Q}$, upto
terms
of $O(\vec{D}^{\bot}/M)$, alongwith the relation between the
normalisation constant $N_{1}=N_{2}$, in the heavy quark mass limit. Here
${\bf\G}$ is some Dirac matrix, possibly with
derivatives. We shall come back to this point at the end of the next section.

We now discuss the consequences of the equality of interpolating fields
for matrix elements, in particular the Bethe-Salpeter amplitude, by
using them in reduction theorems. First of all we would like to address
the question : what does one mean that one is free to choose equivalent
interpolating fields? We will address this question in terms of the
Bethe-Salpeter amplitude but the conclusions apply to any matrix element
involving a heavy meson or baryon.

As an example we consider the heavy vector meson. We define the
Bethe-Salpeter amplitude for the heavy vector meson as
\be
M_{\alpha}\,^{\beta}(x_{1},x_{2})=\langle 0 \vert
\psi_{Q\alpha}(x_{1})G(x_{1},x_{2})\bar{\psi}_{q}^{\beta}(x_{2})
\vert P\rangle\,,\label{bs2}
\ee
In the above equations $\psi_{Q}$ ($\psi_{q}$) is the heavy (light) quark
field and $\vert P\rangle$ represents the heavy vector meson state with a
certain
momentum $P$ and mass $M$. Here time-ordering is implicit.
$G(x_{1},x_{2})$ is a colour matrix chosen to make the B-S
amplitude gauge invariant. However the following analysis also applies
to non gauge invariant amplitudes.

One can now use the reduction theorem to write the B-S amplitude as
\be
M_{\alpha}\,^{\beta}(x_{1},x_{2})=
\lim_{P^{2}\rightarrow M^{2}}(P^{2}-M^{2})\int d^{4}ye^{-iP\cdot y}
\langle 0 \vert
\psi_{Q\alpha}(x_{1})G(x_{1},x_{2})\bar{\psi}_{q}^{\beta}(x_{2})
 \f^{\dagger}(y)\vert 0\rangle\,.\label{bs3}
\ee
The $\f^{\dagger}$ in (\ref{bs3}) is
$\e^{\m} \f^{\dagger}_{\m}$
where $\f_{\m}$ is given in eq. (\ref{ipfv1}) or eq. (\ref{ipfv2}).
Thus one can write
\bea
&&M_{\alpha}\,^{\beta}(x_{1},x_{2})=
\lim_{P^{2}\rightarrow M^{2}}Z_{3}^{1/2}\frac{(P^{2}-M^{2})}{V}\cdot\nonumber
\\
&&\int d^{4}ye^{-iP\cdot y}
\langle 0 \vert
T\psi_{Q\alpha}(x_{1})G(x_{1},x_{2})\bar{\psi}_{q}^{\beta}(x_{2})
\chi_{\g}\,^{\delta}(\partial_{y})\bar{\psi}_{Q}^{\g}(y)\psi_{q\delta}(y)\vert
0\rangle\,,\nonumber\\
&&
\label{bs4}
\eea
where $\chi(\partial_{y})$ is either $\epsslash$ or
$\frac{i\dslash_{y}}{M}\epsslash$.
Correspondingly, $V$ in eq. (\ref{bs4}) is
either $V_{1}$ or $V_{2}$.

The statement that one can use any one of the alternative interpolating
fields means that they should give the same answers in the reduction
theorems. To see what this means consider the second interpolating
field, i.e. $\chi(\partial_{y})=\frac{i\dslash_y}{M}\epsslash$ in the matrix
element (\ref{bs4}). Integrating by parts we can shift the $y$ derivative
in $\chi$ onto
$exp(-iP\cdot y)$ to obtain
\bea
&&M_{\alpha}\,^{\beta}(x_{1},x_{2})=
\lim_{P^{2}\rightarrow
M^{2}}Z_{3}^{1/2}\frac{(P^{2}-M^{2})}{V_{2}}\cdot\nonumber \\
&&\int d^{4}ye^{-iP\cdot y}
\langle 0 \vert
T\psi_{Q\alpha}(x_{1})G(x_{1},x_{2})\bar{\psi}_{q}^{\beta}(x_{2})
(\vslash\epsslash)_{\g}\,^{\delta}\bar{\psi}_{Q}^{\g}(y)\psi_{q\delta}(y)\vert
0\rangle\,.\nonumber\\
&&
\label{bs5}
\eea
With the other choice of the interpolating field we have instead
\bea
&&M_{\alpha}\,^{\beta}(x_{1},x_{2})=
\lim_{P^{2}\rightarrow
M^{2}}Z_{3}^{1/2}\frac{(P^{2}-M^{2})}{V_{1}}\cdot\nonumber \\
&&\int d^{4}ye^{-iP\cdot y}
\langle 0 \vert
T\psi_{Q\alpha}(x_{1})G(x_{1},x_{2})\bar{\psi}_{q}^{\beta}(x_{2})
(\epsslash)_{\g}\,^{\delta}\bar{\psi}_{Q}^{\g}(y)\psi_{q\delta}(y)\vert
0\rangle\,.\nonumber\\
&&
\label{bs5a}
\eea
Thus comparing we see,
that if the two expressions, (\ref{bs5}) and (\ref{bs5a}), are to be equal,
then
$\vslash=\frac{V_{2}}{V_{1}}$, as an operator inside the matrix
element, in the mass shell limit. In general this is not much of a restriction,
as $V_{1}$ and
$V_{2}$ are functions of $v$. However, we have
shown that, in the heavy quark limit, the ratio of the normalisation
constants is unity leading to $\vslash=1$. In fact we see from
(\ref{bs5}) and (\ref{bs5a}) that this condition means that only certain
components of the quark
fields entering in the interpolating fields contribute to the matrix
element, i.e those satisfying
\bea
\bar{\psi}_{Q}(y)\vslash&=&\bar{\psi}_{Q}(y)\nonumber\\
\vslash\psi(y)&=&-\psi(y)\,.
\label{bs5b}
\eea
within the reduction formula, in the mass shell limit.
In other words, in the rest frame of the on-shell {\em meson}, only the quark
part of
the heavy quark field \underline{and} surprisingly also only the antiquark part
of the light quark field
contributes to the matrix element.
   The same relation holds
for the pseudoscalar case. Also the same relation will hold in any matrix
element involving a heavy meson state when we use the interpolating
fields to reduce the state.

It is obvious that this physical requirement will be enforced by taking
the interpolating field for the heavy vector meson, in the heavy quark
limit, to be
\bea
\f_{\m}(x)&=&\frac{1}{2}(\f_{\m}^{1}+\f_{\m}^{2})\nonumber \\
&=&Z_{3}^{1/2}\frac{1}{2V_{1}}[T\bar{\psi}_{q}(x)\g_{\m}^{\bot}\psi_{Q}(x)
-\frac{i\partial_{\n}}{M}T\bar{\psi}_{q}(x)\g^{\n}\g_{\m}^{\bot}\psi_{Q}(x)]\,.
\label{ipfv}
\eea
In fact even if we  start with an arbitrary linear combination of the
two fields, only the sum survives in the heavy quark limit.

With this choice the B-S amplitude picks up the correct projection
operator $\frac{1+\vslash}{2}$ to enforce the above condition and can be
written in the form
\be
M_{\alpha}\,^{\beta}(x_{1},x_{2})=
\chi_{\g}\,^{\delta}(v)A_{\alpha\delta}^{\g\beta}(x_{1},x_{2})\,,\label{bs6}
\ee
with
\be
\chi_{\g}\,^{\delta}(v)=\frac{1}{2}[(1+\vslash)\epsslash]_{\g}\,^{\delta}
\nonumber
\ee
and
\bea
&&A_{\alpha\delta}^{\g\beta}(x_{1},x_{2})=
\lim_{P^{2}\rightarrow
M^{2}}Z_{3}^{1/2}\frac{(P^{2}-M^{2})}{V_{1}}\cdot\nonumber \\
&&\int d^{4}ye^{-iP\cdot y}
\langle 0 \vert
T\psi_{Q\alpha}(x_{1})G(x_{1},x_{2})\bar{\psi}_{q}^{\beta}(x_{2})
\bar{\psi}_{Q}^{\g}(y)\psi_{q\delta}(y)\vert
0\rangle\,.\nonumber\\
&&
\label{bs7}
\eea
The B-S amplitude for the heavy pseudoscalar particle is also of the above
form with $\epsslash$ replaced by $\g_{5}$ and $V_{1}$ by $N_{1}$
\footnote{Although this form of the
B-S amplitude has been forced on us in the heavy quark limit, we could
as well use it for the light meson as after all we are free to choose
any linear combination of the two fields with the appropriate normalisation.}.
It is not very useful as it stands because of the unknown matrix
function $A_{\alpha\delta}^{\g\beta}$. However in the next section we see
that when combined with the HQET
it leads to the previously proposed \cite{hkstw} - \cite{htk} form  of the
heavy meson B-S amplitude.

\section{Bethe-Salpeter Amplitudes and HQET}
\label{hqet}
As is well known by now \cite{kt}, one can go to the zeroth order HQET by
simply replacing the QCD heavy fields in eq. (\ref{bs7}) by
the zeroth order heavy quark effective fields. Thus we shall now simply
consider the heavy quark fields, $\psi_{Q}$, appearing above to be the
corresponding fields appearing in the effective theory.
One can now decouple these heavy fields through the
transformations \cite{kt}
\be
\psi_{Q}(x)=W\left[ {\textstyle{x \atop v}}\right]Q(x)\label{tr1}
\ee
and
\be
\bar{\psi}_{Q}(x)=\bar{Q}(x)W\left[ {\textstyle{x \atop v}}\right]^{-1}\,,
\label{tr2}
\ee
where
\be
W\left[ {\textstyle{x \atop v}}\right]
= P exp\left[
ig\int_{-\infty}^{v\cdot x} dsA\cdot v\right]
\ee
is a path-ordered exponential, wherein the path is a straight line from
$-\infty$ to $x$ along the $v$-direction

Because the transformed fields, $Q$ and $\bar{Q}$ are now decoupled from
the gluons, the right hand side of eq. (\ref{bs6}) factorises and we can
now use the heavy quark propagator
\be
\langle 0\vert Q_{\alpha}(x_{1})\bar{Q}^{\g}(y)\vert 0\rangle
=\int d^{4}p\left[\frac{e^{-ip\cdot (x_{1}-y)}}{\vslash v\cdot p-m_{Q}}
\right]_{\alpha}^{\g}\,\label{hqpr}
\ee
to write the B-S amplitude as
\bea
&&M_{\alpha}\,^{\beta}(x_{1},x_{2})=
\lim_{P^{2}\rightarrow
M^{2}}Z_{3}^{1/2}\frac{(P^{2}-M^{2})}{V_{1}}\cdot\nonumber \\
&&\int d^{4}y d^{4}pe^{-iP\cdot y}
\langle 0 \vert \bar{\psi}_{q}^{\beta}(x_{2})
W\left[ {\textstyle{x_{1} \atop v}}\right]
G(x_{1},x_{2})
W\left[ {\textstyle{y \atop v}}
 \right]^{-1}\left[\frac{e^{-ip\cdot (x_{1}-y)}}{\vslash v\cdot p-m_{Q}}
\right]_{\alpha}^{\g}\chi_{\g}\,^{\delta}(v)\psi_{q\delta}(y)\vert
0\rangle\,.\nonumber\\
&&
\label{bs8}
\eea

The projection operator $\chi(v)$ now gets rid of the $\vslash$ in the heavy
quark propagator to give
\be
M_{\alpha}\,^{\beta}(x_{1},x_{2})=
\chi_{\alpha}\,^{\delta}(v)A_{\delta}\,^{\beta}(x_{1},x_{2})\,,\label{bs9}
\ee
with
\bea
&&A_{\delta}\,^{\beta}(x_{1},x_{2})=
\lim_{P^{2}\rightarrow
M^{2}}Z_{3}^{1/2}\frac{(P^{2}-M^{2})}{V_{1}}\cdot\nonumber\\
&&\int d^{4}y d^{4}pe^{-iP\cdot y}
\frac{e^{-ip\cdot (x_{1}-y)}}{v\cdot p-m_{Q}}\langle 0 \vert
\bar{\psi}_{q}^{\beta}(x_{2})
W\left[ {\textstyle{x_{1} \atop v}}\right]
G(x_{1},x_{2})
W\left[ {\textstyle{y \atop v}}
 \right]^{-1}
\psi_{q\delta}(y)\vert
0\rangle\,.\nonumber\\
&&\label{bs10}
\eea

Eq. (\ref{bs9}) is essentially the result we have been looking for. We can
now
transform this result to momentum space by defining the Fourier
transform
\be
M_{\alpha}\,^{\beta}(p_{1},p_{2})=\chi(v)_{\alpha}\,^{\delta}
A_{\delta}\,^{\beta}(p_{1},p_{2})\,,\label{ftbs1}
\ee
with
\be
A_{\delta}\,^{\beta}(p_{1},p_{2})=
\int d^{4}x_{1}d^{4}x_{2}e^{ip_{1}\cdot x_{1}}e^{ip_{2}\cdot x_{2}}
A_{\delta}\,^{\beta}(x_{1},x_{2})\,.\label{ftbs2}
\ee
The B-S amplitude in this form now satisfies the Bargmann-Wigner
equation on the `heavy' label $\alpha$, leading to all the nice results
of the heavy quark symmetry.

Further, from translation invariance one can write
\be
\langle 0 \vert \bar{\psi}_{q}^{\beta}(x_{2})
W\left[ {\textstyle{x_{1} \atop v}}\right]
G(x_{1},x_{2})
W\left[ {\textstyle{y \atop v}}
 \right]^{-1}
\psi_{q\delta}(y)\vert 0\rangle=\int d^{4}qd^{4}ke^{-iq\cdot (x_{2}-x_{1})}
e^{-ik\cdot (x_{1}-y)}\Delta_{\delta}\,^{\beta}(q,k)\,,\label{gf1}\,
\ee
leading to
\be
A_{\delta}\,^{\beta}(p_{1},p_{2})=
\lim_{P^{2}\rightarrow M^{2}}Z_{3}^{1/2}\frac{(P^{2}-M^{2})}{V_{1}}
\delta^{4}(P-p_{1}-p_{2})\int
d^{4}k\frac{\Delta_{\delta}\,^{\beta}(p_{2},k)}
{v \cdot (P-k)-m_{Q}}\,.\label{ftbs3}
\ee
One way to generate the physical pole at
$P^{2}=M^{2}$ is
for the Green's function $\Delta(p_{2},k)$ to be peaked at
$v\cdot k=\Lambda$, where $\Lambda=(M-m_{Q})$. In other words the
``longitudinal" mass of the
light degrees of freedom is $\Lambda$. This is of course trivially true in the
weak
binding or equal velocity approximation, where there is no transverse
momentum and $m_{q}=\Lambda$.

It is now easy to generalise to arbitrary orbital resonances. One can
take as interpolating fields, for orbital angular momentum $L$, either
\be
\f_{L}^{1}(x)=Z_{3}^{1/2}\frac{1}{L_{1}}T\bar{\psi}_{q}(x){\bf\G}\psi_{Q}(x)
\label{ipfl1}
\ee
or
\be
\f_{L}^{2}(x)
=-Z_{3}^{1/2}\frac{1}{L_{2}}\frac{i\partial_{\n}}{M}
T\bar{\psi}_{q}(x)\g^{\n}{\bf\G}\psi_{Q}(x)
\label{ipfl2}
\ee
where the ${\bf\G}$ are listed in the table for the four possible spin-parity
states. We have omitted Lorentz indices and the $L_{i}$ are
normalisation constants. Again using the equations of motion it is easy to show
that the
difference between these two possible interpolating fields is always
either of the form
$\bar{\psi}_{q}\cdots\frac{\stackrel{\leftrightarrow}{\partial}^{\bot}\cdot
\vec{D}^{\bot}}{m_{Q}+m_{q}}\psi_{Q}$ or
$\bar{\psi}_{q}\cdots\frac{\vec{D}_{\m}^{\bot}}{m_{Q}+m_{q}}\psi_{Q}$. As usual
these
are small in the heavy quark limit. (Here
$\stackrel{\leftrightarrow}{\partial}_{\m}=\vec{\partial}_{\m}
-\stackrel{\leftarrow}{\partial}_{\m}$.) Hence in the heavy quark mass
limit we find $\phi^{1}_{L}(x)=\phi^{2}_{L}(x)$ and $L_{1}=L_{2}$.
Thus the rest of the analysis follows in general. One can see that it is
also, in principle, quite easy to generalise to heavy baryons.

\section{Acknowledgements}
We would like to thank John Strathdee for very useful discussions.

\section*{Table Caption.}
Lowest gamma structure \cite{htk} appearing in interpolating fields for mesonic
L-wave states.

\begin{table}[htbp]
\centering
\begin{tabular}{ccl} \hline\hline
\bf state & $\bf J^{PC}$&$ {\bf \G}$ \\ \hline
&& \\
${}^{3}L_{L-1}$ &$ (L-1)^{(-)^{L+1}(-)^{L+1}}$ &
$\stackrel{\leftrightarrow}{\dslash}^{\bot}
\stackrel{\leftrightarrow}{\partial}_{\m_{2}}^{\bot}\ldots
\stackrel{\leftrightarrow}{\partial}_{\m_{L}}^{\bot} $ \\
&& \\
${}^{3}L_{L}$ &$ L^{(-)^{L+1}(-)^{L+1}}$ &
$\g_{5}[\stackrel{\leftrightarrow}{\dslash}^{\bot},\g_{\m_{1}}^{\bot}]
\stackrel{\leftrightarrow}{\partial}_{\m_{2}}^{\bot}\ldots
\stackrel{\leftrightarrow}{\partial}_{\m_{L}}^{\bot} $ \\
&& \\
${}^{3}L_{L+1}$ &$ (L+1)^{(-)^{L+1}(-)^{L+1}}$ &
$\g_{\n}^{\bot}\stackrel{\leftrightarrow}{\partial}_{\m_{1}}^{\bot}\ldots
\stackrel{\leftrightarrow}{\partial}_{\m_{L}}^{\bot} $ \\
&& \\
${}^{1}L_{L}$ &$ L^{(-)^{L+1}(-)^{L}}$ &
$\g_{5}\stackrel{\leftrightarrow}{\partial}_{\m_{1}}^{\bot}\ldots
\stackrel{\leftrightarrow}{\partial}_{\m_{L}}^{\bot} $\\
&& \\ \hline
\end{tabular}
\end{table}
\end{document}